\documentclass[10pt,conference]{IEEEtran}

 \usepackage{amsmath,amssymb,amsfonts}
 \usepackage{textcomp}
\usepackage{xspace}
\usepackage{listings}
\usepackage[normalem]{ulem} %emphasize still italic
\usepackage{array, booktabs}
\usepackage{blindtext}
\usepackage{ifthen}

\usepackage{csvsimple,supertabular}
\usepackage[most]{tcolorbox}
\usepackage[caption=false]{subfig}
\usepackage{paralist} %to list extends of previous work
\usepackage{parskip}
\usepackage{tabularx}
\usepackage{fontawesome}      % for icon next to scg URL next to author list on title page
\usepackage{adjustbox} 		% for vertical text in tables
\usepackage{pifont} 		% for crosses and ticks in tables
\newcolumntype{R}[2]{
    >{\adjustbox{angle=#1,lap=\width-(#2),margin*=0.4em 0em 0em 0em}\bgroup}
    l
    <{\egroup}
}
\let\oldFootnote\footnote
\newcommand\nextToken\relax
\renewcommand\footnote[1]{%
    \oldFootnote{#1}\futurelet\nextToken\isFootnote}
\newcommand\isFootnote{%
    \ifx\footnote\nextToken\textsuperscript{,}\fi}

\newcommand{\id}[1]{$-$Id: scgPaper.tex 32478 2010-04-29 09:11:32Z oscar $-$}

\newcommand{\ie}{\emph{i.e.},\xspace}
\newcommand{\eg}{\emph{e.g.},\xspace}
\newcommand{\etal}{\emph{et al.}\xspace}
\newcommand{\etc}{\emph{etc.}\xspace}

\newcommand{\SO}{{SO}\xspace}

\newcommand{\ml}{{mailing lists}\xspace}
\newcommand{\SOid}[1]{[\href{https://www.stackoverflow.com/questions/#1}{\emph{SO:#1}}]}

\newcommand{\smallfootnote}[1]{\footnote{\fontsize{6}{6}\selectfont #1}}

\newcolumntype{L}[1]{>{\raggedright\arraybackslash}m{#1}} % left aligned column

\def\BibTeX{{\rm B\kern-.05em{\sc i\kern-.025em b}\kern-.08em
	T\kern-.1667em\lower.7ex\hbox{E}\kern-.125emX}}
	
\newboolean{showedits}
\setboolean{showedits}{true} % toggle to show or hide edits
\ifthenelse{\boolean{showedits}}
{
	\newcommand{\del}[1]{\textcolor{red}{\sout{#1}}} % please delete
	\newcommand{\nbe}[3]{
		{\colorbox{#3}{\bfseries\sffamily\scriptsize\textcolor{white}{#1}}}
		{\textcolor{#3}{\sf\small$\blacktriangleright$\textit{#2}$\blacktriangleleft$}}}
}{
	\newcommand{\del}[1]{} % please delete
	
	\newcommand{\nbe}[3]{}
}

\ifthenelse{\boolean{showedits}}
{
	\newtcolorbox{inserted}{
		title=Inserted text:,
		colframe=blue,colback=blue!5!white,
		breakable,
		leftrule=0mm, 
		bottomrule=0mm,
		rightrule=0mm,
		toprule=0mm,
		arc=0mm, outer arc=0mm,
		oversize
	}
	\newtcolorbox{deleted}{
		title=Deleted text:,
		colframe=red,colback=red!5!white,
		breakable,
		leftrule=0mm, 
		bottomrule=0mm,
		rightrule=0mm,
		toprule=0mm,
		arc=0mm, outer arc=0mm,
		oversize
	}
	\newtcolorbox{refactored}{
		title=Rewritten text:,
		colframe=blue,colback=red!5!white,
		breakable,
		leftrule=0mm, 
		bottomrule=0mm,
		rightrule=0mm,
		toprule=0mm,
		arc=0mm, outer arc=0mm,
		oversize
	}
}{

}

\newboolean{showcomments}
\setboolean{showcomments}{true}
\ifthenelse{\boolean{showcomments}}
{\newcommand{\nbc}[3]{
		{\colorbox{#3}{\bfseries\sffamily\scriptsize\textcolor{white}{#1}}}
		{\textcolor{#3}{\sf\small$\blacktriangleright$\textit{#2}$\blacktriangleleft$}}}
	}
{\newcommand{\nbc}[3]{}
	}

\definecolor{source}{gray}{0.9}
\newboolean{isblinded}
\setboolean{isblinded}{true}
\ifthenelse{\boolean{isblinded}}
{\newcommand\blind[1]{BLINDED\xspace}}
{\newcommand\blind[1]{#1\xspace}}

\newcommand{\seclabel}[1]{\label{sec:#1}}
\newcommand{\secref}[1]{\autoref{sec:#1}}
\newcommand{\figlabel}[1]{\label{fig:#1}}
\newcommand{\figref}[1]{\autoref{fig:#1}}
\newcommand{\tablabel}[1]{\label{tab:#1}}
\newcommand{\tabref}[1]{\autoref{tab:#1}}

\newcommand{\lstref}[1]{\autoref{lst:#1}}

\usepackage[pdftex,colorlinks=true,pdfstartview=FitV,linkcolor=black,citecolor=black,urlcolor=black,bookmarks=false]{hyperref}

\newcommand{\rqI}{What \textbf{high-level topics} do developers discuss about code comments?}
\newcommand{\rqII}{What \textbf{type of questions} do developers ask about code comments?}
\newcommand{\rqIII}{What \textbf{information needs} do developers seek about commenting practices?}
\newcommand{\rqIV}{What specific commenting conventions are recommended by developers?}

\newcommand{\repFolder}[1]{Folder ``\href{https://doi.org/10.5281/zenodo.3762776}{RP/#1}'' in the Replication package}
\newcommand{\repFile}[1]{File ``\href{https://doi.org/10.5281/zenodo.3762776}{RP/#1}'' in the Replication package}
\newcommand{\makar}{Makar\xspace}
\begin{document}

\title{What Do Developers Discuss about Code Comments?}

\author{
\IEEEauthorblockN{
Pooja Rani\IEEEauthorrefmark{1},
Mathias Birrer\IEEEauthorrefmark{1}, 
Sebastiano Panichella\IEEEauthorrefmark{2},  
Mohammad Ghafari\IEEEauthorrefmark{3},
Oscar Nierstrasz\IEEEauthorrefmark{1}}\\
\IEEEauthorblockA{\IEEEauthorrefmark{1}Software Composition Group, University of Bern\\Bern, Switzerland\\\href{http://scg.unibe.ch/staff/}{\faGlobe \hspace{0.1cm}scg.unibe.ch/staff}}
\IEEEauthorblockA{\IEEEauthorrefmark{2} Zurich University of Applied Science (ZHAW), Switzerland\\
    \href{mailto:panc@zhaw.ch}{panc@zhaw.ch}}
    \IEEEauthorblockA{\IEEEauthorrefmark{3} University of Auckland, New Zealand\\
    \href{mailto:m.ghafari@auckland.ac.nz}{m.ghafari@auckland.ac.nz}}
}

\maketitle
\begin{abstract}
Code comments are important for program comprehension, development, and maintenance tasks. 
Given the varying standards for code comments, and their unstructured or semi-structured nature, developers get easily confused (especially novice developers) about which convention(s) to follow, or what tools to use while writing code documentation. 
Thus, they post related questions on external online sources to seek better commenting practices. 
In this paper, we analyze code comment discussions on online sources such as Stack Overflow (\SO) and Quora to shed some light on the questions developers ask about commenting practices.
We apply Latent Dirichlet Allocation (LDA) to identify emerging topics concerning code comments.
Then we manually analyze a statistically significant sample set of posts
to derive a taxonomy that provides an overview of the developer questions about commenting practices.\\
Our results highlight that on \SO nearly 40\% of the questions mention how to write or process comments in documentation tools and environments, and nearly 20\% of the questions are about potential limitations and possibilities of documentation tools to add automatically and consistently more information in comments.
On the other hand, on Quora, developer questions focus more on background information (35\% of the questions) or asking opinions (16\% of the questions) about code comments.\\
We found that 
(i) not all aspects of comments are covered in coding style guidelines, \eg how to add a specific type of information,
(ii) developers need support in learning the syntax and format conventions to add various types of information in comments, and
(iii) developers are interested in various automated strategies for comments such as detection of bad comments, or verify comment style automatically, but lack tool support to do that.
\end{abstract}


\section{Introduction}
%Developers use software documentation to understand software codebase~\cite{Bavo13b}.
%Developers have access to several types of documentation, for example, wikis, user manuals, and code comments.
%However, they find code comments and source code more trustworthy than other forms of documentation~\cite{Maal14a}.
%Recent studies provide evidence that developers consider code comments to be the most important type of documentation for understanding code~\cite{Souz05a}.
%Studies show that code comments make the code comprehension easier compared to uncommented code~\cite{Wood81a,Tenn85a}.
%For this reason, code comments are not only useful for code comprehension but also for software maintenance tasks~\cite{Wood81a,Hart93a,Souz06a,Corn09a}.
Recent studies provide evidence that developers consider code comments to be the most important type of documentation for understanding code~\cite{Souz05a}.
Code comments are written using natural language sentences, and their syntax is neither imposed by a programming language's grammar nor checked by its compiler.
Consequently, developers follow various conventions in writing code comments~\cite{Padi09a}.
These conventions vary across development environments as developers embed different kinds of information in different environments~\cite{Pasc17a,Zhan18a,Rani21b}.
This makes it hard to write, evaluate, and maintain the quality of comments (especially for new developers) as the software evolves~\cite{Alla14a,Kern99a}.

To help developers in writing readable, consistent, and maintainable comments,  programming language communities, and large organizations, such as Google and Apache Software Foundation provide coding style guidelines that also include comment conventions \cite{Java20a,Stru00a,Goog20a,Spar20a}.
%These style guidelines cover various aspects of comment conventions such as syntactic, stylistic, and content-related conventions.
%Whereas, syntactic refers the syntax to write different kinds of comments (class/method/package comment),stylistic refers to the format, writing style and grammatical style of comments,  and content-related refers to the type of information should be written in the comments.
%For example, \emph{``{Use 3rd person (descriptive) not 2nd person (prescriptive)}''} is an example of a stylistic comment convention for Java documentation comments.\smallfootnote{\url{https://www.oracle.com/technical-resources/articles/java/javadoc-tool.html}, accessed on 10 Aug, 2020}
%In addition to explicitly mentioned guidelines, unspoken common practices often arise in projects when developers create and adopt personalized style guidelines for the languages.\smallfootnote{\url{https://github.com/povilasb/style-guides/blob/master/cpp.rst}, accessed on 10 Aug, 2020}
%These common practices are hard to turn into strict rules and do not have enough tool support to verify them.
However, the availability of multiple syntactic alternatives, the freedom to adopt personalized style guidelines,\smallfootnote{\url{https://github.com/povilasb/style-guides/blob/master/cpp.rst}, accessed on Jun, 2021} and the lack of tools for assessing comments, make developers confused about which commenting practice to adopt~\cite{Alla14a}, or how to use a tool to write and verify comments.  
%These conventions vary across different development environments such as the documentation tool, or the IDEs (integrated Development Environment) used.
%Thus, developers often gets confused about the commenting practices to apply.
%Miltiadis \etal defined this as the \emph{convention inference problem}~\cite{Alla14a}.

To resolve potential confusion, and to learn best commenting practices, developers post questions on various Q\&A forums.
Stack Overflow (\SO) is one of the most popular Q\&A forums, enabling developers to ask questions to experts and other developers.\smallfootnote{\url{https://www.stackoverflow.com}}
%Barua \etal measures the share of a topic (\emph{topic share}) as the proportion of posts that contain the topic and discovered that the ``coding style'' topic is the one with the highest \emph{topic share} value on \SO~\cite{Baru14a}.
Barua \etal determined the relative popularity of a topic across all \SO posts and discovered the ``coding style'' topic as the most popular~\cite{Baru14a}.
Similarly, Quora\smallfootnote{\url{https://www.quora.com}} is another widely adopted by developers to discuss software development aspects \cite{Krug19a}.
However, what specific problems developers report about code comments such as do they face challenges due to multiple writing conventions or development environments, or which commenting conventions experts recommend to them on these sources, is unknown.

Therefore, we analyze commenting practices discussions on \SO and Quora, to shed light on these concerns.
Particularly, we formulate the following research questions:
\begin{enumerate}
\item \emph{\textbf{RQ$_1$}: \rqI} Our interest is to identify high-level concerns and themes developers discuss about code comments on Q\&A platforms.
\item \emph{\textbf{RQ$_2$}: \rqII} Our aim is to identify the type of questions developers frequently ask \eg questions such as how to write comments, or what is the problem in their comments. In addition, we aim to identify which platform they prefer to ask which type of questions.
\item \emph{\textbf{RQ$_3$}: \rqIII} We investigate \SO and Quora questions in more detail (including body, tags, comments of the question) to identify the challenges and needs related to writing comments in various development environments. 
\item \emph{\textbf{RQ$_4$}: \rqIV} We investigate the answers to specific kinds of questions, asking about best practices, to collect commenting conventions suggested by developers.
\end{enumerate}

For each research question, we analyze developer questions at various levels, such as focusing only on the title of the question, the whole question body, or the answers to the question.
% such as focusing only on the title of the question to obtain the main intent, the whole body of the question to understand their needs, or the answers for the question to study the recommendations.
%, for example, \emph{RQ$_1$} aims at a general overview of all SO questions while the remaining questions focus on a sample of selected questions.
%Specifically, \emph{RQ$_2$} analyzes only the question title to know the main intent, \emph{RQ$_3$} analyzes the whole question body to better understand their needs, and \emph{RQ$_4$} focuses on the answers recommended by other developers.
The rationale behind each level is that future approaches for identifying and automating developers' intent, needs, and recommendations can focus on that specific aspect of comments they want to evaluate and improve.
Our manually labeled questions for the research questions \emph{RQ$_2$}, \emph{RQ$_3$}, and \emph{RQ$_4$} can serve as an initial dataset for building such approaches.

To answer \textbf{RQ$_1$} we use a semi-automated approach involving LDA~\cite{Blei03a}, a well-known topic modeling technique used in software engineering, to explore topics from the \SO and Quora posts~\cite{Baru14a,Wang13b,Yang16a,Pokh19a}.
We then analyze a statistically significant sample set of posts from \SO and Quora posts and derive a taxonomy of \emph{types of questions} and \emph{information needs} of developers regarding comments to answer \textbf{RQ$_2$} and \textbf{RQ$_3$} respectively.
To answer \textbf{RQ$_4$}, we manually analyze the questions asked about best practices on the selected sources and extract various commenting conventions recommended by developers in their answers.
%The taxonomy offers an overview of the leading developer questions discussing commenting conventions, in a more formal, structured, and possibly exhaustive way.

Our results show that developers frequently ask questions on Q\&A forums to discuss the best syntactic conventions to write comments, ways to retrieve comments from the code or background information about various comment conventions. 
Specifically, the questions about how to write or process comments (\emph{implementation strategies}) in a language, tool, or IDE  are frequent on \SO. 
On the other hand, questions about background information concerning various conventions or opinions on the best commenting practices are frequently posted on Quora.
Our analysis shows that developers are interested in embedding various kinds of information, such as code examples and media (\eg images), in their code comments but lack the strategies and standards to write them.
%As organizations enforce their own coding style guidelines in addition to the standard programming language guideline, developers face difficulties in locating particular guidelines to write various kinds of information.
%In case of absence of such guidelines, developers post questions about the best practices to write code comments.
%We found such questions in 15\% of all questions .
%We find 15\% of all questions are about comprehending the best practices to write code comments.
%This indicates the need for standardizing documentation guidelines and assuring their findability for developers.
%Tomasottir \etal showed in his interview study that developers use automatic style checkers (linters) to learn programming language conventions, maintain code consistency and save discussion time~\cite{Toma17a}.
%Therefore we encourage novice developers to use automatic style checkers to fasten their learning process about code comments. 
We also observe a considerable proportion of questions where developers ask about the ways of automating the commenting workflow with documentation tools or IDE features to foster commenting practices and assess them.
This shows the increasing need to improve the state of commenting tools by emphasizing better documentation of the supported features and by providing their seamless integration in the development environments.
%This indicates the need to improve the findability of documentation guidelines, and tools to write and verify the comments. 
%Additionally, with the increasing use of multi-language projects, there is a need of flexible documentation tools and automated style checkers that allow developers to customize the default configuration of these tools and integrate them into the development pipeline better.
%On Quora, we observed that only 13\% of commenting questions are relevant to code comments and most of these questions correspond to opinions on how to add comments in the code, and practices for writing different information types in the comments.

The contributions of this paper are:
\begin{enumerate}
\item an empirically-validated taxonomy of comment-related concerns collected from multiple sources;
\item a first study to investigate the Quora platform for code comments; and 
\item a publicly available dataset including all validated data, and steps to reproduce the study in the Replication Package (RP)~\cite{Scam21a}.
\end{enumerate}

\textbf{Paper structure.} 
In \secref{study-design} we detail the study definition and methodology adopted to answer our research questions. 
In \secref{results}, we present our results and insights.
We discuss our findings and their implications in \secref{discussion}.
We recap the threats to validity in \secref{threats-to-validity}, summarize related work in \secref{related-work}, and conclude the paper in \autoref{sec:conclusion}.

%===============================================================================
%=============================== STUDY DESIGN ===================================
%===============================================================================
\section{Study Design}
\seclabel{study-design}

The goal of this study is to investigate information needs, practices, and problems developers discuss on various online Q\&A platforms about code comments.  
\figref{study-pipeline} illustrates the steps followed to answer the research questions.

\begin{figure}[ht]
\center
\includegraphics[width=0.99\linewidth]{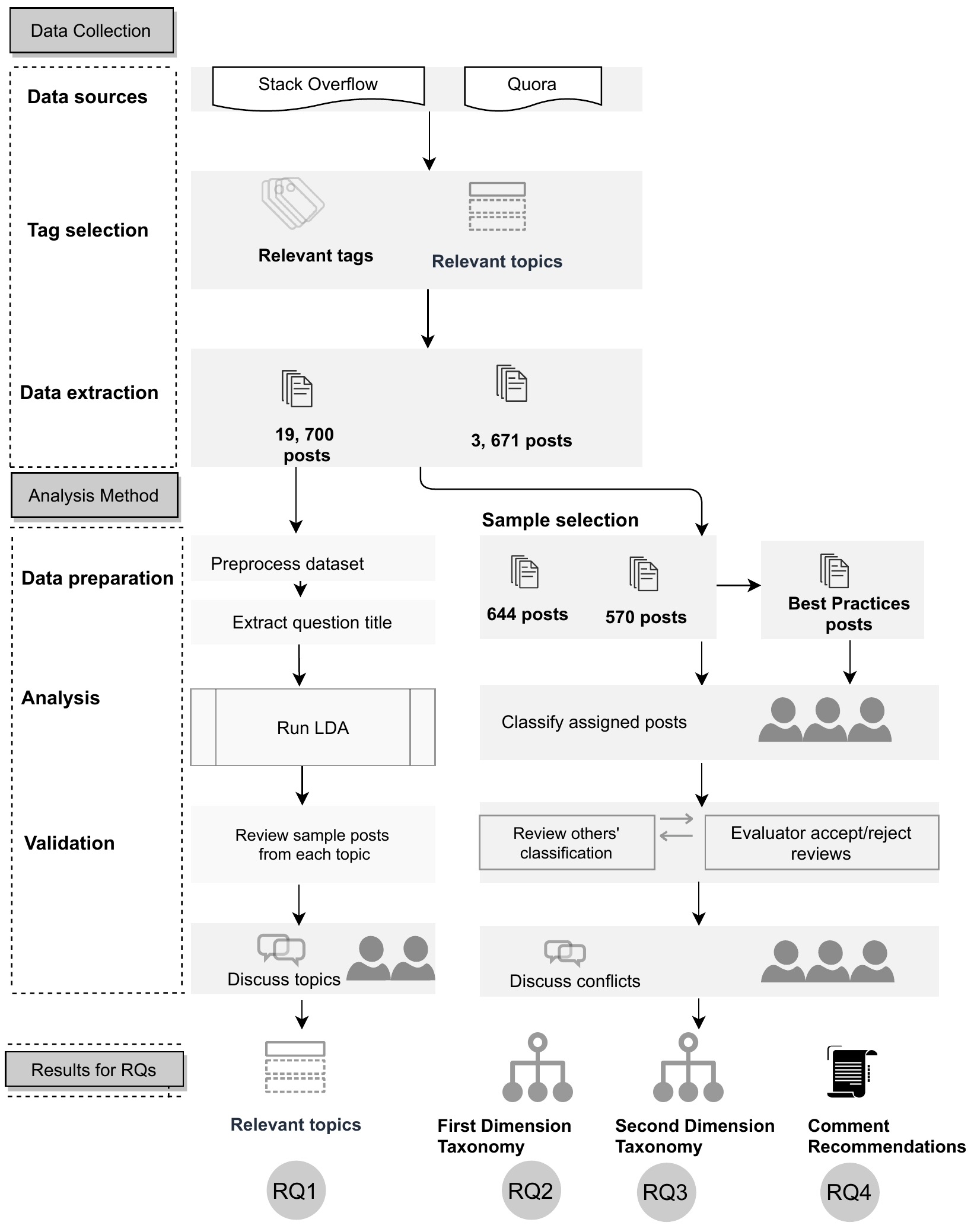}
\caption{	Research Approach to answer research questions}
\figlabel{study-pipeline}
\vspace{-2mm}
\end{figure}

\subsection{Data collection}
\label{subsec:data-collection}
\textbf{\SO.} 
%We collected the data from \SO using the \emph{Stack Exchange Data Explorer} interface.\smallfootnote{\url{https://data.stackexchange.com/}}
%The interface facilitates the users to query all Stack Exchange sites in a SQL-like Query language. 
%First of all, we extracted the questions having at least one answer. 
To identify the relevant discussions concerning \emph{commenting practices},
 we used an approach similar to Aghajani~\cite{Agha19a}.
We selected the initial keywords ($Ik$) such as \emph{comment}, \emph{convention}, and \emph{doc} to search the \SO tags page~\cite{Agha19a}.\smallfootnote{\url{https://stackoverflow.com/tags}  verified on Jun 2021}
The search converged to a set of 70 potentially relevant tags, referred to as \emph{initial tags} ($It$).

Two authors independently examined all the tags, their descriptions and the top ten questions in each tag, and selected the relevant tags.
%This examination of 26 tags resulted in ten tags, by discarding the unrelated or duplicate tags.
% about the development frameworks (which provide various features to attach comments in the websites), the tags about code element conventions, and about plugin conventions.
%In this process, we discarded irrelevant tags, such as the \emph{django-comment} tag, which refers to the Django comment framework to enable users to add comments to a website.
We observed that certain tags are ambiguous due to their usage in a different context, such as the \emph{comments} tag being used in many settings.
For example, the tag \emph{comment} (5\,710 questions on \SO) contains questions
about development frameworks (\eg Django) that provide the feature of attaching comments to a website or other external source (478 questions tagged with ``wordpress''), or about the websites where users add comments to the posts (512 questions tagged with ``facebook'' tag).
%For example, developers frequently post questions with the tag \emph{comment} (5\,710 questions on \SO) but not all the questions are related to code comments.
%The questions can be about development frameworks (\eg Django) that provide the feature of attaching comments to a website or other external source (478 questions tagged with ``wordpress''), or about the websites where users add comments to the posts (512 questions tagged with ``facebook'' tag).
Therefore, we discarded posts where co-appearing tags were \emph{wordpress} and \emph{facebook}, or \emph{django-comment}.
%Similarly,  the tag \emph{facebook-comments} is about accessing the comments section of Facebook programmatically.
%In another instance, the tags \emph{code-convention} and \emph{coding-conventions} from the \emph{convention} keyword redirect to \emph{coding-style} tag page.
%Another author reviewed the list of initial tags and agreed on them or discussed them mutually with the previous author in case of disagreement to discard some irrelevant tags.
The resulting set of tags ($It$) is therefore reduced to 55 tags out of 70 tags.
The list of selected 55 tags is available in the RP.\smallfootnote{\repFile{Tags-topics.md}}

We extracted all questions tagged with at least one tag from the $It$ set, resulting in 19\,705 non-duplicate questions.
For each question, we extracted various metadata fields such as ID, title, body, tags, creation date, and view count from \SO using the \emph{Stack Exchange Data Explorer} interface.\smallfootnote{\url{https://data.stackexchange.com/}}
The interface facilitates the users to query all Stack Exchange sites in an SQL-like Query language.\smallfootnote{\repFile{Stack-exchange-query.md}}
%Aghajani \etal studied software documentation-related posts, including code comments on \SO and other sources~\cite{Agha19a}.
%We extracted their documentation-related tags from the given replication package and compared them to our tags ($It$) to verify if we missed any.

\begin{comment}
\begin{table}[h]
\centering
\caption{Extracted fields from \SO and Quora}
\tablabel{so-quora-fields}
\vspace{-2mm}
\begin{tabular}{ p{0.15\linewidth}lp{0.5\linewidth} }
\noalign{\smallskip}\hline
\textbf{Source} & \textbf{Field} & \textbf{Description} \\
\noalign{\smallskip}\hline
\SO 
& ViewCount & Number of views for that question \\
& CreationDate & Creation date of the question \\
& Id & Question ID \\
& Title & Question title\\
& Body & Question body\\
& Tags & Question tags\\
\hline\noalign{\smallskip}
Quora & url &  Question URL \\
& title & Question title\\
& body &  Question body\\
& topics & Topics tagged in the question\\
& answers & All answers of the question\\
\noalign{\smallskip}\hline
\end{tabular}
\end{table}
\end{comment}

\textbf{Quora.}
Extracting data from Quora was non-trivial due to the lack of publicly available datasets and services to access the data,  its restrictive scraping policies~\cite{Pati16a}, and the absence of a public API to access the data. 
Thus, to extract its data, we implemented a web scraper in Python using \emph{selenium} to automate browsing, and \emph{BeautifulSoup} to parse the HTML.\smallfootnote{[\url{https://pypi.org/project/beautifulsoup4/}], [\url{https://www.selenium.dev/}]}
On Quora, the notion of topics is the same as tags on \SO, so a question in Quora can be tagged with topics similar to \SO tags.
Unlike the \SO tag page, Quora provides neither an index page listing all its topics, nor a list of similar topics on a topic page.
% ON: PLEASE don't add useless commas -- "We therefore" doesn't need commas.
We therefore used the relevant \SO tags as initial Quora topics,  searched for them on the Quora topic search interface, and
obtained 29 topics, such as \emph{Code Comments}, \emph{Source Code}, \emph{Coding Style}. The list of all topics and their mapping to \SO tags is provided in the RP.\smallfootnote{\repFile{Tags-topics.md}}
We scraped all questions with their meta-data such as URL, title, body, and topics of each question from the identified topics, resulting in 3\,671 questions in total.
%Hence, for each question, we extracted the meta-data shown in \tabref{so-quora-fields}.
%To validate the questions' relevancy and mitigate one author's bias, another author inspected the irrelevant questions and found three disagreement cases. 
%The disagreement cases were inspected by the third author who has not yet seen the posts and resolved using a majority voting mechanism.

\subsection{Analysis Method}
\seclabel{methodology}
%To explore developer high-level concerns, we applied LDA-based topic modeling approach to find out high-level topics about code comments.
%Then we manually analyzed a statistically significant sample set of questions from \SO and Quora to derive the taxonomy and gather the comment conventions recommended by developers.
%To perform the manual analysis task, we designed a \emph{three-iterations based process} involving three authors. 
%In the first iteration, we classified the allotted posts, as shown in the step \emph{Classify assigned posts} in \figref{study-pipeline}.
%In the second iteration, all authors involved validated each other's classification to mitigate potential biases and check possible incorrect clarifications shown in \emph{Validation} step in \figref{study-pipeline}.
%In the end, all authors resolved all disagreements by discussing conflicts and topics using a majority voting mechanism.
%The detailed methodology can be visualized in \figref{study-pipeline} and described in the following parts.

\textbf{Automated analysis from LDA (RQ$_1$)}.
LDA infers latent discussion topics to describe text-based documents. Each document can contain several topics, and each topic can span several documents, thus making it possible for the LDA model to discover ideas and themes in a corpus.
We applied LDA on the \SO dataset but excluded the Quora dataset as it contains a high number of irrelevant posts (nearly 80\%) based on the manually analyzed statistically significant sample set shown in \tabref{various-sources-data}.
Additionally, as LDA uses the word frequencies and co-occurrence frequencies across documents to build a topic model of related words, having a high number of irrelevant posts can impact the model quality.
Since our objective is to discover the high-level concerns developers have, we extract only titles of the \SO questions, as the title summarizes the main concern while the body of the question adds non-relevant information, such as details of development environment, what the developer has tried, or sources already referred to.

To achieve reliable high-level topics from LDA, we performed the following data-preprocessing steps on the question titles:
removal of HTML tags, code elements, punctuation and stop words (using the Snowball stop word list\smallfootnote{\url{http://snowball.tartarus.org/algorithms/english/stop.txt}}), and applied Snowball stemming\cite{Chen16a}.
We used the data management tool, \makar, to prepare the data for LDA~\cite{Maka20a}.
We provide the concrete steps \makar performed to preprocess the data in the \emph{Reproducibility} section in the online appendix~\cite{Scam21a}.
The preprocessed \emph{title} field of the questions served as the input documents for LDA.
We used the \emph{Topic Modeling Tool}~\cite{Jona17a}, a GUI for MALLET~\cite{McCal} that uses a Gibbs sampling algorithm, and facilitates extending the results with meta-data.
We provide the input data\smallfootnote{\repFolder{RQ1/LDA\_input}} used for the MALLET tool and the output\smallfootnote{\repFolder{RQ1/LDA\_output}} achieved in the RP.

LDA requires optimal values for the $k, \alpha$, and $\beta$ parameters to be chosen, which depends on the type of data under analysis, but this represents an open challenge in software engineering tasks.
Wallach \etal pointed out that choosing a smaller \emph{k} may not separate topics precisely, whereas a larger \emph{k} does not significantly vary the quality of the generated topics~\cite{Wall09a}.
Therefore, to extract distinct topics that are both broad and high-level, we experimented with several values of \emph{k} ranging from 5 to 25, as suggested by Linares-Vásquez \etal~\cite{Lina13a}.
%For example, comment syntax, IDEs \& editors, databases, and documentation generation are high-level topics related to comments but distinct from each other.
%The goal was to find a wide enough range of high-level topics which were still clearly distinct, employing the most relevant keywords and meaning. 
We assessed the optimal value of \emph{k} by analyzing the topic distribution, coherence value (large negative values indicate that words do not co-occur together often)\smallfootnote{\url{http://mallet.cs.umass.edu/diagnostics.php}} \cite{Rode15a}, and perplexity score (a low value means the model correctly predicts unseen words) \cite{Hoff10a} for each value of \emph{k} from the given range~\cite{Chan09b}.
This process suggested $k=[10]$ as the most promising value for our data (with the lowest perplexity of -6.9 and high coherence score of -662.1) as fewer redundant topics were selected with these values.

In the next iterations, we optimized the hyperparameters $\alpha$ and $\beta$ by using the best average probability of assigning a dominant topic to a question, inspired by the existing studies~\cite{Rose16a}.
%~\cite{Rose16a,Pani13a}.
We selected the initial values of hyperparameters $\alpha=\frac{50}{k} \beta=0.01$ using the de facto standard heuristics~\cite{Bigg14a} but allowed these values to be optimized by having some topics be more prominent than others.
We ran the model optimizing after every ten iterations in total 1000 iterations.
Thus, we concluded that the best hyperparameter configuration for our study is $k=10, \alpha=5, \beta=0.01$.
%In addition to above configuration, we ran LDA for 1000 iterations, optimizing every ten iterations with random seed value equal to 100.
%This configuration resulted in an average topic probability for the dominant topic of $72\%$.
%According to the configuration, we ran LDA and collected the topic results where each topic presents a set of words arranged according to their likelihood of belonging to the topic.
% but without a meaningful topic name.
As LDA does not assign meaningful names to the topics, we manually inspected a sample of 15 top-ranked questions under each topic to assign topic names.
%Based on the inspections, we identified some semantically similar topics in terms of related words and questions asked.
%For example, topic six \emph{Naming convention in project} and topic seven \emph{Naming code entities} shown in \tabref{topics-lda} are distinct from each other in word order but similar to each other in terms of context of the words as both topics focus on naming conventions of various code entities thus can be merged if required.
%For such similar topics, we investigated a sample of 20 top-ranked questions to confirm their similarity.

\begin{table}[htp]
	\centering
	\caption{Data extracted from \SO and Quora sources}
	\tablabel{various-sources-data}
	\vspace{-2mm}
	\begin{tabular}{p{0.25\linewidth}p{0.17\linewidth}p{0.17\linewidth}p{0.17\linewidth}}
	\noalign{\smallskip}\hline
	\textbf{Source} &  \textbf{Extracted posts} &  \textbf{Manually analyzed} &  \textbf{Relevant posts}\\
	\hline
	\SO & 19\,700 & 644 & 416 \\
	Quora & 3\,671 & 565 & 118\\
	\noalign{\smallskip}\hline
	\end{tabular}
\end{table}

\textbf{Taxonomy Study (RQ$_2$, RQ$_3$).} After extracting all posts tagged with the relevant tags for \SO, we analyzed a statistically significant sample from all (19\,700) posts, reaching a confidence level of 99\% and an error margin of 5\%.
The resulting sample set contains 644 posts.
We selected the sample posts using a random sampling approach without replacement to reach 644 posts.
Similarly, for Quora, we selected 565 posts for our manual analysis, as shown in \tabref{various-sources-data}.

\emph{{Classification:}}
We classified the sample posts into a \emph{two-dimensional taxonomy} mapping concepts of the selected questions.
The first dimension (\emph{question types}) aims to answer RQ$_2$ while the second dimension (\emph{information needs}) answers RQ$_3$.
The first dimension, inspired from an earlier \SO study~\cite{Beye14a}, defines the categories concerning the kind of question, \eg if a developer is asking how to do something related to comments, what is the problem with their comments, or why comments are written in a particular way.
%The first dimension (\emph{question types}) categorizes the questions according to the categories similar to the one reported in prior studies~\cite{Alla13a,Rose16a}.
We renamed their categories~\cite{Beye14a} to fit our context, \eg their `What' type of question renamed to `Implementation problems.'
To classify the selected sample questions in the first dimension categories as shown in \tabref{first-dimension-categories}, we used closed card sorting technique.

%To get more insight into developer information needs such as they ask question about which specific development environment (\eg which programming languages, tool, IDEs), and on which platforms they ask these questions more frequently, we categorize the sampled questions based on the second dimensions.
The second dimension outlines more finely-grained categories about the types of information needs developers seek~\cite{Guzz13a}, \eg development environment-related needs (\eg comments in programming languages, tools, IDEs), or about comments in general.
The majority of these categories are built based on the software documentation work by Aghajani \etal \cite{Agha19a}, and the questions are classified into these categories using the hybrid card sorting technique~\cite{Mill12a}.
In the development environment-related needs, we identified if a question talks about \emph{IDE \& Editors} (\eg Intellij, Eclipse), \emph{Programming languages} (Java, Python), or \emph{Documentation tools} (Javadoc, Doxygen).
The further sub-levels of the taxonomy focus on the type of information a questioner is seeking in each development environment, such as asking about the syntax to add a comment or specific information in the comment in Javadoc~\cite{Guzz13a}.
%For example, if the question is asking about syntax \& format to add comments or a specific type of information in them, understand or process comments for specific development tasks, or adapting the comment templates for their tasks.
For instance, the question \emph{``How to reference an indexer member of a class in C\# comments''}~\cite{SO:379346} is about the C\# language, and asking about the syntax to refer to a member in the class comment, thus gets classified into the three levels as \emph{Programming languages}|\emph{Syntax \& format}|\emph{Class comment} according to the taxonomy shown in \figref{so-quora-full-dimension-categories}.
%In addition, we also classify if a user inquires about an existing feature or existence of a feature (asking for a feature). This way, we can highlight the area where documentation tools can improve.

\begin{table*}[htp]
\centering
\small
\caption{First dimension categories}
\tablabel{first-dimension-categories}
\vspace{-1mm}
%\begin{adjustbox}{width=\linewidth}  
%\begin{tabular}{p{0.15\linewidth}p{0.58\linewidth}p{0.29\linewidth}}
		     \begin{center}
       \resizebox{0.95\linewidth}{!}{
       \scriptsize	

\begin{tabular}{p{0.15\linewidth}p{0.50\linewidth}p{0.35\linewidth}}
\noalign{\smallskip}\hline
\textbf{Category} & \textbf{Description} & \textbf{Question keywords to identify with an example} \\
\noalign{\smallskip}\hline
{Implementation Strategies} & The questioner is not aware of ways to write or process comments. 
They often ask questions about integrating different information in their comment, using features of various tools.
& ``How to'', \eg \emph{How to use @value tag in javadoc?} \\
\noalign{\smallskip}\hline
{Implementation Problems} &  The questioner tried writing or processing the code comment but was unsuccessful.
& the question ``What is the problem?'', \eg \emph{Doxygen \textbackslash{command} does not work, but @command does?}\\
\noalign{\smallskip}\hline
{Error} & The questioner posted the error, exceptions or crashes while writing or generating comments, or any warning produced by the documentation tool. & contain an error message from the exceptions or stack trace \\
\noalign{\smallskip}\hline
{Limitation \& Possibilities} & The questioner is seeking more information about limitations of a comment related approach, tool, or IDE, and various possibilities to customize the comment. & the question ``is it possible or allowed'', \eg \emph{Is there a key binding for block comments in Xcode4?} \\
\noalign{\smallskip}\hline
{Background Information} &  The questioner is looking for background details on the behavior of comments in a programming languages, a tool, or a framework. & the question ``why something'', \eg \emph{Why in interpreted languages the \# usually introduces a comment?} \\
\noalign{\smallskip}\hline
{Best practice} & The questioner is interested to know the best practice, guidelines or general advice to tackle a comment-related problem or convention. & the question ``is there a better way to'', \eg \emph{What is the proper way to reference a user interaction in Android comments?}\\
\noalign{\smallskip}\hline
{Opinion} & The questioner is interested to know the judgment of other users for a comment convention. & the question ``what do you think'', \eg \emph{Are comments in code a good or bad thing?}\\
\hline
\end{tabular}
	}
\end{center}
%\end{adjustbox}
\vspace{-3mm}
\end{table*}

\emph{Execution and Validation:} 
A Ph.D. candidate, a master's student, and a faculty member, each having more than three years of programming experience, participated in the evaluation of the study.
The sample set of questions was divided into an equal subset of questions and selected random questions for each subset to ensure that each evaluator gets a chance to look at all types of questions.
We followed a three-iteration-based approach to categorize the questions.
In the first iteration, we classified the posts into first and second-dimension categories.
In the second iteration, each evaluator (as a reviewer) reviewed the classified questions of other evaluators and marked their agreement or disagreement with the classification.
In the third iteration, the evaluator agreed or disagreed with the decision and changes proposed by the reviewers.
In case of disagreements, another reviewer who had not yet looked at the classification reviewed the classification and gave his/her decision.
Finally, if all evaluators disagreed, we chose the category based on the majority voting mechanism.
This way, it was possible to ensure that each classification is reviewed by at least one other evaluator.
In case of questions belonging to more than one category, we reviewed the other details of questions, such as tags and comments of the questions and chose the most appropriate one.
We finalized the categories and their names based on the majority voting mechanism.
%One example for such a scenario is given in the \textbf{RQ$_2$} subsection of \secref{results}.
%To validate the question classification and category name, we used a two-step validation approach.
%In the first step, two reviewers validated the assigned category of the question as described above.
%In the next step, all evaluators independently validated the category name and then discussed the conflicts among themselves to reach the consensus.

Based on the classification and validation approach described above, all three authors (evaluators) evaluated first the relevance of all their assigned questions, and reviewed the cases of irrelevant questions marked by other evaluators.
The third author reviewed and resolved their disagreement cases using a majority voting mechanism (cohen's k = 0.80).
As a result, 416 questions of \SO and 118 questions of Quora were considered relevant to our study, as shown in \tabref{various-sources-data}.
The remaining questions, marked as irrelevant, were manually inspected and no new relevant topic was identified.

\textbf{Recommended comment conventions (RQ$_4$).}
Given the unstructured or semi-structured nature of comments, and varying standards to write comments, various organizations and language communities present numerous commenting guidelines to support consistency and readability of the comments.
For instance,  the convention ``\emph{Use 3rd person (descriptive) not 2nd person} in writing comments'' is given in the Java Oracle style guide.
However, not all of these conventions are recommended by developers in real-time and some conventions are even discouraged, depending on the development environment.
Additionally, developers assumed some conventions were feasible \eg overriding docstrings of a parent class in its subclasses, but other developers pointed out them as a limitation of current documentation tools or environment.
We attempted to collect such comment conventions recommended by developers in their answers on \SO and Quora.
From the classified questions in RQ$_2$, we chose the questions categorized in the \emph{Best Practice} category according to \tabref{first-dimension-categories}. 
Based on the accepted answers of these questions, we identified \emph{recommendation} or \emph{limitation} of various comment conventions.
In case a question has no accepted answer, we referred to the top-voted answer.

\section{Results}
\seclabel{results}
%===============================================================================
%=============================== LDA results ======================================
%===============================================================================
%\vspace{-2mm}
\subsection{High-Level Topics Discussed about Comments (RQ$_1$)}

\begin{table*}[ht]
	\centering
	\footnotesize
	\small
	\caption{Topics generated by LDA with assigned topic name and ten most important topic keywords}
	\vspace{-2mm}
	\tablabel{topics-lda}
	
		     \begin{center}
       \resizebox{0.95\linewidth}{!}{
       \scriptsize
	\begin{tabular}{p{0.06\linewidth}p{0.16\linewidth}p{0.70\linewidth}}
	\noalign{\smallskip}\hline
	 \textbf{Relevance} & \textbf{Topic name} & \textbf{Topic words} \\
	\noalign{\smallskip}\hline
	 R & Syntax \& format & line code file c python doxygen block php html text remov string tag javascript style add regex command script singl \\ 
	 \noalign{\smallskip}\hline
	 R & IDEs \& Editors & javadoc generat studio eclips visual file xml java project code class c android sourc maven tag doc intellij show netbean \\ 
	 \noalign{\smallskip}\hline
	 R & R documentation & r rmarkdown markdown tabl output pdf html file knitr code render chunk text packag chang latex error knit plot add \\ 
	\noalign{\smallskip}\hline
	 R & Code conventions & function class method jsdoc type doxygen paramet c object variabl return python phpdoc javadoc name refer properti convent valu docstr \\ 
	 \noalign{\smallskip}\hline
	 IR & Development frameworks for thread commenting & api doc rest rail generat spring rubi test net swagger web rspec where asp creat find develop googl rdoc what \\ 
	 \noalign{\smallskip}\hline
	 IR & Open source software & code sourc what open can app where whi get find anyon websit develop program if android mean softwar doe someon \\ 
	 \noalign{\smallskip}\hline
	 R & Documentation generation & sphinx file doxygen generat python html link doc page includ modul make creat build custom autodoc rst restructuredtext imag output \\ 
	\noalign{\smallskip}\hline
	 IR & Thread comments in websites & facebook post wordpress page get php whi plugin user box like show youtub section display form system repli delet add \\ 
	 \noalign{\smallskip}\hline
	 IR & Naming conventions \& data types & convent name what java python develop c whi sql tabl valu case mysql string data x program code variabl column \\ 
	\noalign{\smallskip}\hline
	 R & Seeking documentation \& learning language & what code software best program way write good language develop standard requir tool project c practic need learn are which \\ 
	\noalign{\smallskip}\hline
	\end{tabular}
	}
\end{center}
	\vspace{-2mm}
	\end{table*} 
	
\tabref{topics-lda} shows the 10 topics generated from the LDA analysis, where the column \emph{Relevance} denotes if a topic is relevant (R), or irrelevant (IR) and  \emph{Topic name} is the assigned topic label.
The column \emph{Topic words} shows the words generated by LDA, sorted in the order of their likelihood of relevance to the topic. 
%Among all 10 topics identified by LDA and reported in \tabref{topics-lda}, we classified six of them as being relevant for code comment discussions. 

%The topic \emph{Open source} reports questions about developing and publishing open-source applications or software. Majority of these questions were posted on Quora.
In the {\emph{Syntax \& Format}} topic, developers mainly ask about the syntax of adding comments, removing comments, parsing comments, or regex to retrieve comments from code.
Occasionally, the questions are about extracting a particular type of information from comments to provide customized information to their clients, such as obtaining descriptions, or to-do comments from code comments.
Depending on a programming language or a tool, strategies to add information in the comments vary, such as adding a description in XML comments~\SOid{9594322}, in the R environment~\SOid{45917501}, or in the Ruby environment~\SOid{37612124}.
This confirms the relevance of recent research efforts on identifying the comment information types in various programming languages~\cite{Pasc17a,Zhan18a,Rani21d}.
\emph{IDEs \& Editors} groups questions about commenting features provided in various IDEs to add or remove comments in the code, or setting up documentation tools to write comments.
\emph{R Documentation} groups questions about documentation features provided in the R language, such as creating various formats of documents including Markdown, Latex, PDF, or HTML documentation.
`R Markdown', a documentation format available in knitr package, provides these features in R.
In the \emph{Code convention} topic groups the questions about best practices, such as printing the docstrings of all functions of an imported module, or conventions to check types in Javascript.
In this topic, developers also ask the reasons behind various code conventions such as the reason behind having only one class containing the main method in Java, or using particular symbols for comments.
In the \emph{Documentation generation} topic,  
developers inquire about problems in generating project documentation or HTML documentation from comments automatically using various documentation tools such as Sphinx, Doxygen.
%, or how can they document their code in a way that facilitates the automatic generation of HTML documentation from the code comments.
Apart from code comments, software projects support other forms of documentation, such as wikis, user manuals, API documentation, or design documentation.
As the project documentation is divided into various components, developers post questions about locating them in the topic {\emph{Seeking documentation}}.
Additionally, developers also showed interest in learning various programming languages and thus seeking documentation to learn them.
%Additionally, developers seek help in validating this information in the comments \eg validating Python comments~\SOid{36554318}.
%Overall, developers complains about not able to locate various syntax or format related conventions or guides to use or integrate the documentation tools in their development environment.
%Developers are also found to be interested in automating their code documentation and thus ask for various features for it.

\noindent\fbox{%
	\parbox{\linewidth}{%
		\fontsize{9}{10}\selectfont
		\textbf{Finding 4.1.1} 
		The most relevant topics discussed by developers about commenting practices, and identified by LDA are related to  
		\emph{Syntax \& format}, \emph{Documentation generation}, \emph{IDEs \& Editors}, \emph{R documentation}, \emph{Code conventions}, and \emph{Seeking documentation}, indicating developers interest in automating their code documentation.
	}%
}
\vspace{1mm}

\subsection{Types of Questions Discussed ({RQ$_2$})} 
To get insights about which types of questions developers ask about comments and where they ask the specific types of questions more often, we categorized the sampled set of questions from \SO and Quora according to the first dimension (\emph{question type}) shown in \tabref{first-dimension-categories}.
\figref{first-second-dimension-categories} shows such categories on the x-axis with respect to both sources, and the y-axis indicates the percentage of questions belonging to a category out of the total question of a source. 
The figure highlights \emph{implementation strategies} (how-to) to be the most frequent category on \SO, confirming prior study results~\cite{Beye19a,Alla13a,Wang13b,Baru14a}.
Differently from previous studies, we found \emph{best practice} and \emph{background information} questions to arise more frequently than \emph{implementation problems} questions.

\begin{figure}[t]
\centering
\includegraphics[width=\linewidth]{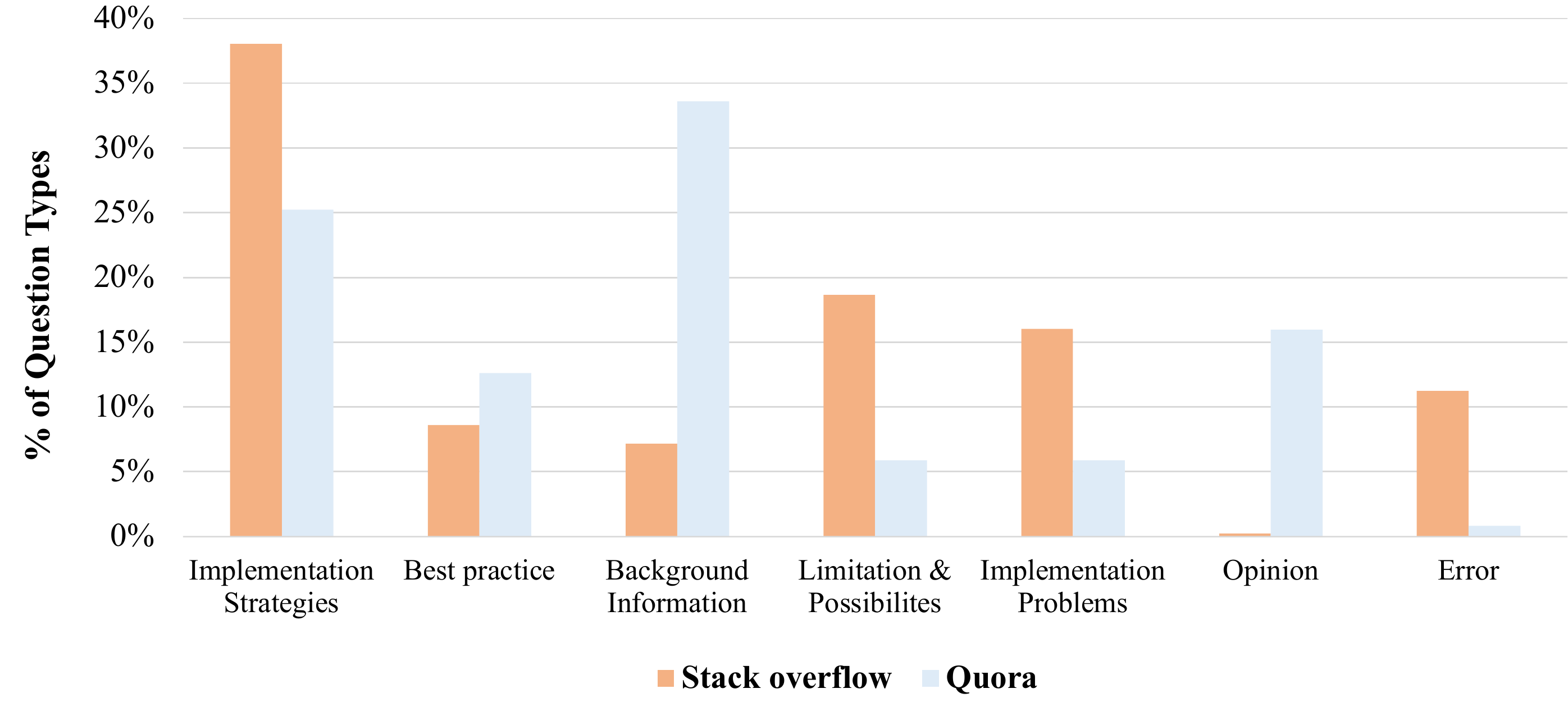}
\caption{First dimension categories found on \SO and Quora}
\figlabel{so-quora-first-dimension-categories}
\end{figure}
%We calculate for each source the ratio of the number of posts belonging to a category to the total number of posts in the source.
We also observed that different types of questions are prevalent on the investigated platforms, as highlighted by \figref{so-quora-first-dimension-categories}.
The figure shows that developers ask \emph{implementation strategies} questions and \emph{implementation problems} questions more on \SO compared to Quora.
Despite Quora being an opinion-based Q\&A site, we also observed questions about \emph{best practice} and \emph{background information} about reasons behind various implementation and symbols used in comments.
This shows how developers rely on Quora to gather knowledge behind numerous conventions and features provided by the development environment.
Such types of questions are also found on \SO but to a lesser extent. 
For instance, we observed the question: \emph{What are these tags @ivar @param and @type in python docstring}~\SOid{379346} on \SO.
Based on the thousands of views count of the post, we can say that the question has attracted the attention of many developers.
Uddin \etal gathered developers' perceptions about APIs in \SO by mining opinions automatically \cite{Uddi19a}. 
Our study provides the evidence to include Quora as another source to validate their approach and mine developer's opinions.

\vspace{1mm}
\noindent\fbox{%
	\parbox{\linewidth}{%
		\fontsize{9}{10}\selectfont
	\textbf{Finding 4.2.1} 
Different kinds of questions are prevalent on \SO and Quora \eg \emph{implementation strategies} and \emph{implementation problems} questions are more common on \SO whereas \emph{best practice} and \emph{background information} questions apart from opinion-based questions are more prevalent on Quora.
This suggests that Quora can be a useful resource to understand how developers perceive certain development aspects, while \SO is useful to understand what technical challenges they face during its development.x
	}%
}

\vspace{1mm}
\subsection{Developer Information Needs (\textbf{RQ$_3$})} 
\begin{figure}[t]
	\centering
	\includegraphics[width=\linewidth]{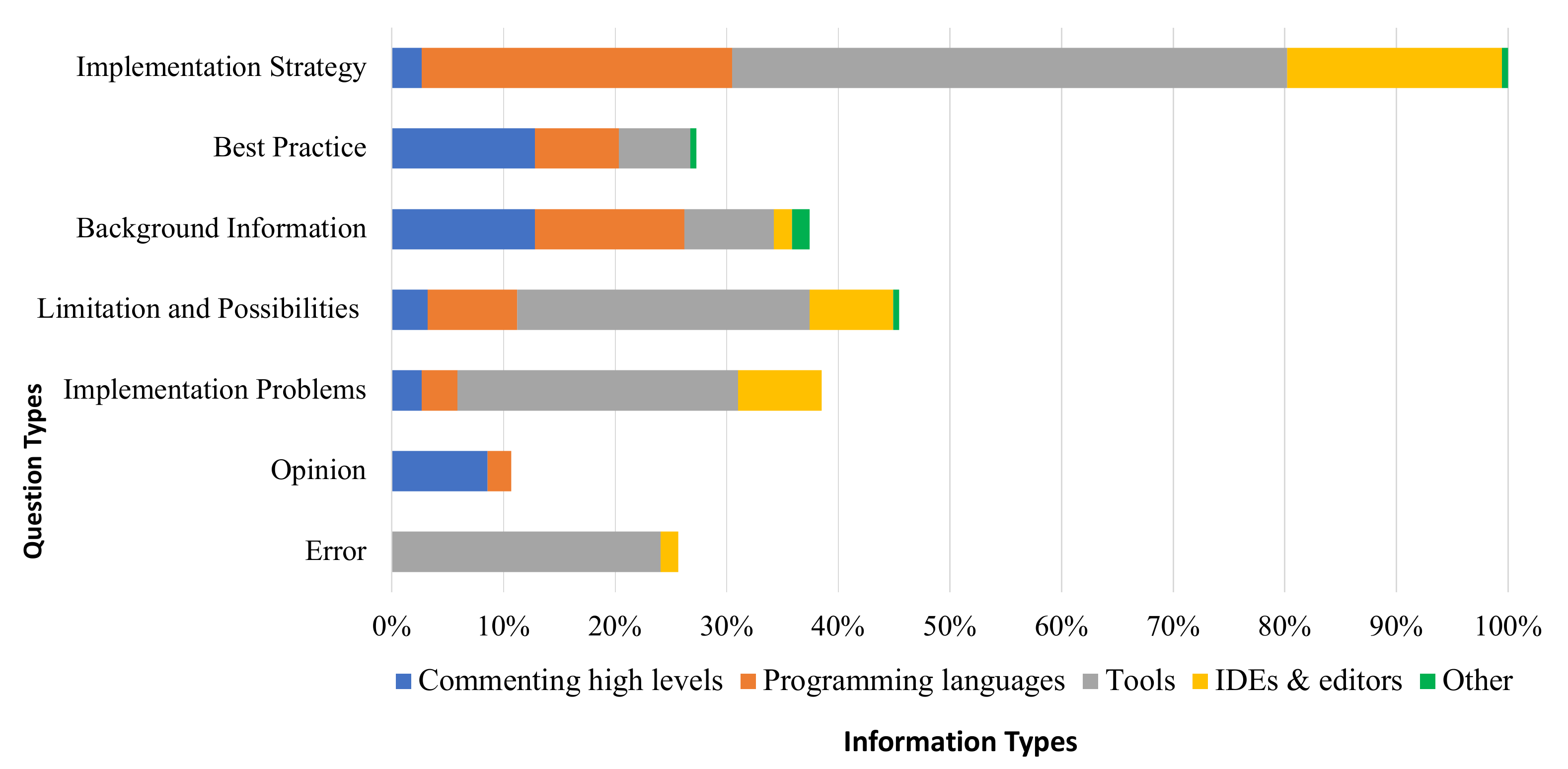}
	\caption{Distribution of first dimension and second dimension categories}
	\figlabel{first-second-dimension-categories}
	\vspace{-3mm}
\end{figure}
We analyzed the questions from two different perspectives.
The first-dimension categories (\emph{question types}) (Y-axis in \figref{first-second-dimension-categories}) show the types of question (\eg implement strategies, problems) developers ask, whereas the second-dimension categories (X-axis in \figref{first-second-dimension-categories}) highlight the kinds of problems they face with the code comments in the development environment.
For example, \figref{first-second-dimension-categories} shows how in the second dimension analysis the most frequent category \emph{implementation strategies} contains questions about how to do something (comment-related) in a development environment, be it specific to a programming language, a documentation tool, or to the IDE itself.
On the other hand, developers discuss the possible features of the documentation tools and IDEs in the \emph{limitation \& possibilities} category.
This category highlights the developer's struggle in locating the feature details from the documentation of such tools and showcases the vital need for improving this aspect.
However, which specific features and syntaxes of comments developers seek in the development environment is essential information to progress in this direction.
Therefore, we first separated general questions about comment conventions to the \emph{commenting high levels} category, and moved other development environment related questions to the \emph{programming languages}, \emph{tools}, and \emph{IDEs \& editors} categories (first-level categories shown in \figref{so-quora-full-dimension-categories}).
We then added sub-categories such as \emph{Syntax \& format}, \emph{Asking for feature}, \emph{Change comment template} \etc under each first-level category to highlight the specific need related to it, as shown in \figref{so-quora-full-dimension-categories} and explained in Table III~\cite{Scam21a}.
In the next paragraphs, we explain a few such subcategories.
%We observed that this way one-third of the questions are about the development environment. 
%In these questions, the top two frequent categories are \emph{Syntax \& format} and \emph{Asking for feature}.

\vspace{1mm}
\noindent\fbox{%x
\parbox{\linewidth}{%
\fontsize{9}{10}\selectfont
\textbf{Finding 4.2.2} 
One-third of the developer commenting practices questions are about their specific development environment. The top two frequent questions concern the categories \emph{Syntax \& format} and \emph{Asking for feature} indicating developers' interest in improving their comment quality.
The rest focus on setting up or using documentation tools in IDEs to generate comments automatically.
}%
}\\

\begin{figure}[ht]
    \centering
    \includegraphics[width=.9\linewidth]{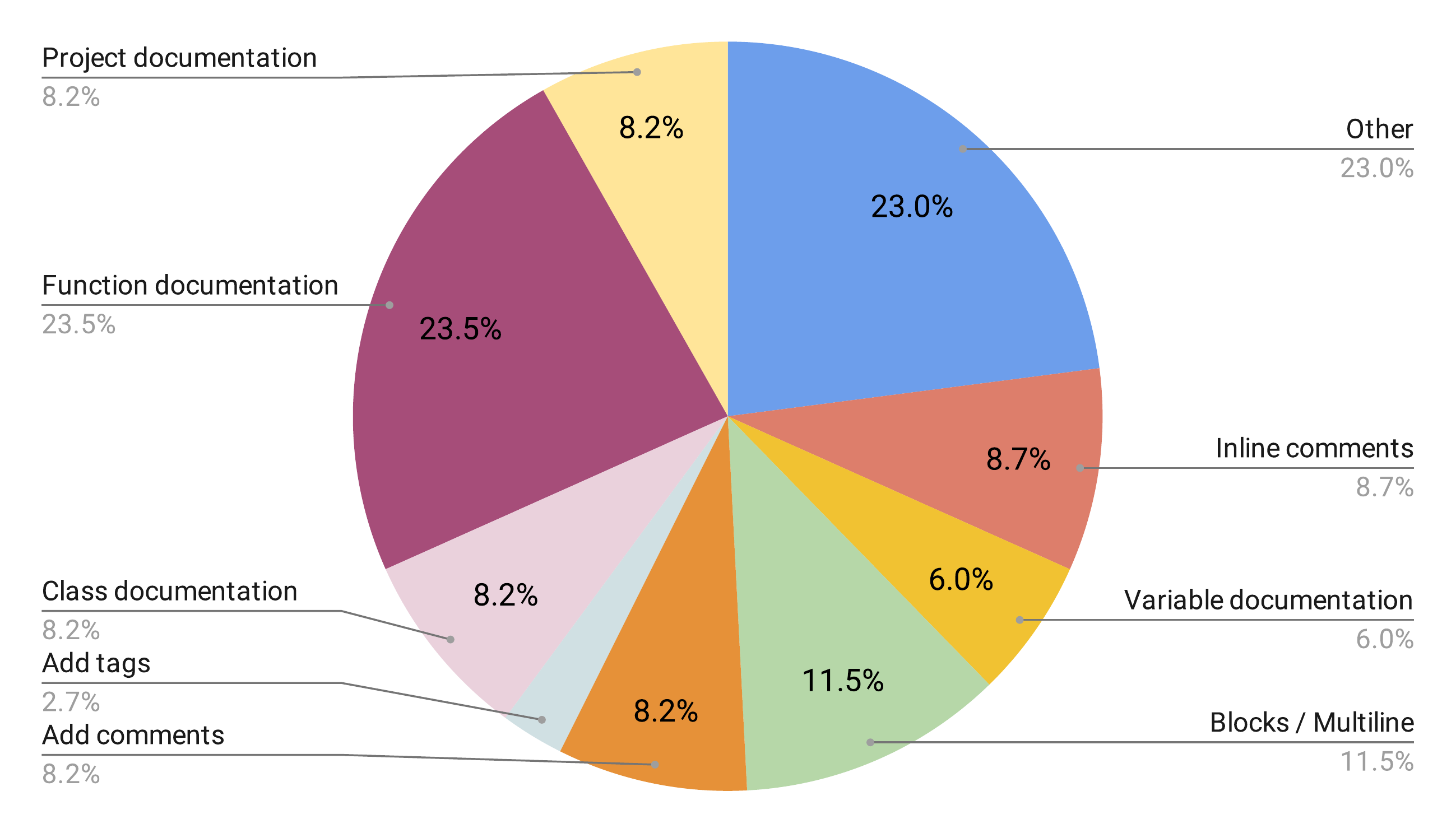}
    \caption{Comment's syntax \& format discussions}
    \figlabel{syntax-format}
    \vspace{-2mm}
\end{figure}

\begin{figure}[ht]
    \centering
    \includegraphics[width=\linewidth]{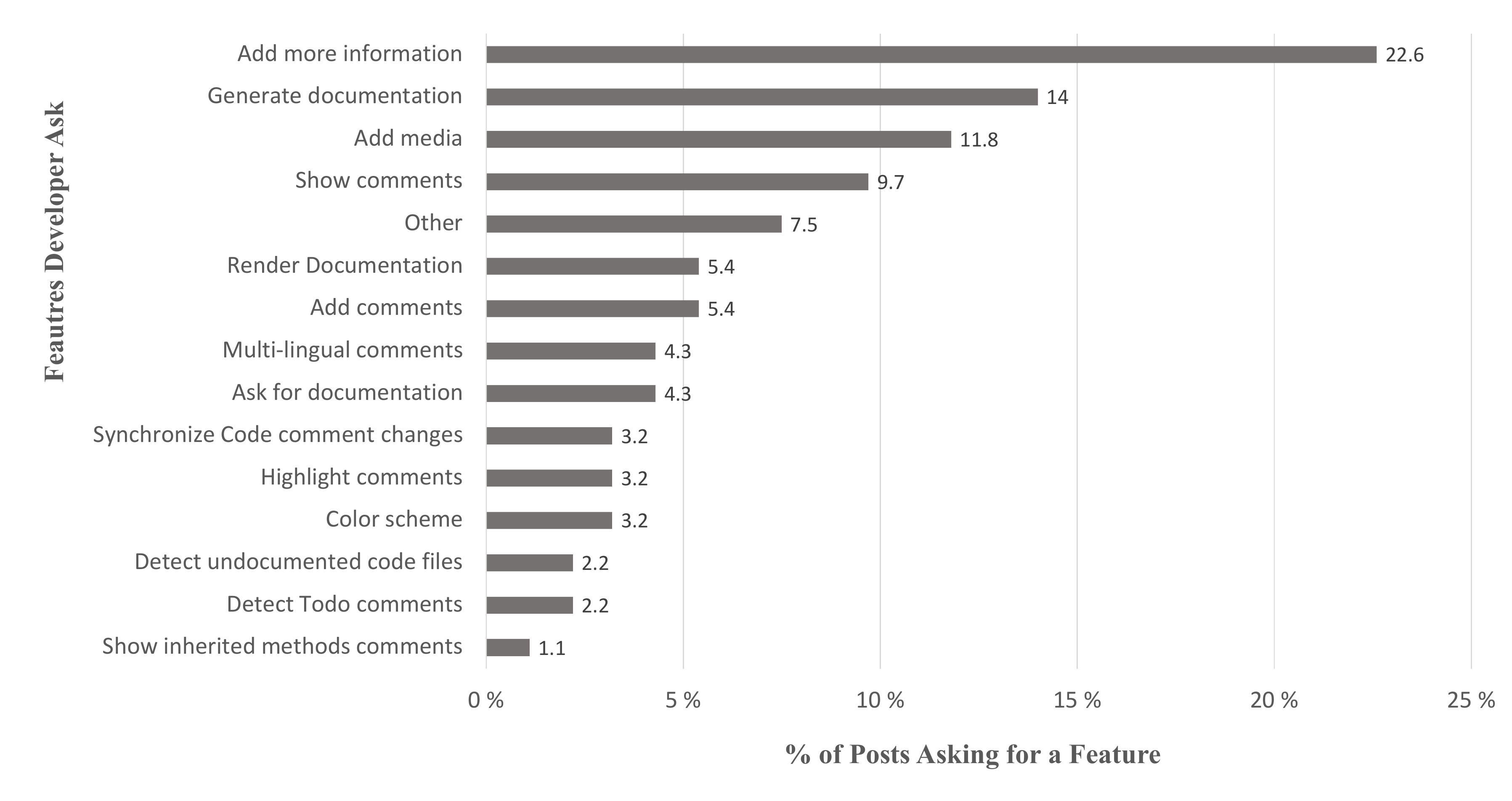}
    \caption{Developers asking for features}
    \figlabel{asking-for-feature}
        \vspace{-2mm}
\end{figure}

In \emph{Syntax \& format}, shown in \figref{syntax-format}, developers discuss syntax to add various kinds of comments to the code, such as function comments, class comments, block comments, and different tags.
%183 posts (so and quora) belongs to syntax and format
Specifically, the syntax of writing function comments is asked more frequently (23\% of questions) than other types of comments, showing the importance and efforts of API documentation.
In this analysis we found 23\% of the questions marked as \emph{Other} are either about the syntax of writing comments in a programming language or a tool without stating the kind of comment (class/function/block), or concerning the intent of syntax conventions. 
Such background questions are more often posted on Quora compared to \SO. 

\emph{Asking for feature} is another frequent information developers seek on \SO to locate various features provided by the documentation tools.
We rarely found such questions on Quora.
Aghajani \etal reported a similar category as \emph{Support/Expectations} covering developer needs that are not satisfied by the documentation tools in their work~\cite{Agha19a}.
In our study context, we reported such inquiries, as shown in \figref{asking-for-feature}, and in the category \emph{Asking for feature} under all development environment categories in \figref{so-quora-full-dimension-categories}.

\figref{asking-for-feature} shows that developers frequently need to add different kinds of information to the code comments, such as code examples: \emph{How to add code description that are not comments?}~\SOid{45510056}, performance-related: \emph{Javadoc tag for performance considerations}~\SOid{39278635}, and media~\SOid{43556442}.
Additionally, developers ask about features to add comments automatically, detect various information from the comments, or synchronize the comments with code changes~\SOid{23493932}.
These questions show the worthiness of devoting research efforts to the direction of identifying information types from comments,
%~\cite{Pasc17a,Zhan18a},
 detecting inconsistent comments,
% ~\cite{Wen19a,Rato17a},
and
assessing and generating comments automatically
%~\cite{Wang19a,Srid10a} 
to improve code comments~\cite{Pasc17a,Wen19a}.
We separated the feature-related questions (different features of the tools and IDEs) into two categories, \emph{Using feature} and \emph{Asking for feature}, based on the user awareness.
In the former category, the user is aware of the existence of a feature in the environment but finds problems in using it, as shown in \lstref{tool-using-feature-valueTag}.
In the latter category, users inquire about the existence of a feature, or try to locate it, as shown in \lstref{tool-asking-feature-addInlineImages}.

\begin{minipage}{0.9\linewidth}\centering
\begin{lstlisting}[caption= {Using @value feature in Javadoc}, label={lst:tool-using-feature-valueTag}]
How to use @value tag in javadoc?
\end{lstlisting}
\end{minipage}

\begin{minipage}{0.9\linewidth}\centering
\begin{lstlisting}[caption= {Asking for a feature to add inline images}, label={lst:tool-asking-feature-addInlineImages}]
How can I show pictures of keyboard keys in-line with text with Sphinx?
\end{lstlisting}
\end{minipage}

\vspace{1mm}
\noindent\fbox{%
\parbox{\linewidth}{%
\fontsize{9}{10}\selectfont
\textbf{Finding 4.2.3} 
The \emph{implementation strategies} category questions are the most frequently viewed questions on \SO and \emph{limitation \& possibilities} (is it possible) questions are the second most viewed questions based on the view count of questions.
On the other hand, based on the answer count of questions, \emph{best practice} questions trigger the most discussions along with \emph{implementation strategies}.
}%
}

%Such questions on the Q\&A sites hint at the lack of such information in the tool's documentation or expectations raised from other tools'' experience that provide these features.
%A similar instance is shown in the question \emph{Bilingual (English and Portuguese) documentation in an R package}~\SOid{37288823}, where a developer asks a question about the possibility of adding comments in two languages to show it depending on the language region.
%The experts inform the developer about the non-existence of such a feature in the R programming language.
%Besides, they direct the developer to other ways of achieving a similar result.
%We argue that such questions do not only provide instances of the unique case scenarios but also provide an opportunity to tool designers to improve their tool or at least improve their documentation by better listing and explaining the supported (and unsupported) features.
In addition to the above categories, we observed that \SO encourages developers, especially novice developers, to ask questions about the basics of various topics~\cite{Koch16a}, grouped into the \emph{commenting high levels} category, and shown in \figref{so-quora-full-dimension-categories}.
This detailed taxonomy of the second dimension is reported in the \figref{so-quora-full-dimension-categories} and Table III in the online appendix~\cite{Scam21a}.
%Table III shows the hierarchy of categories constructed in the second dimension \ie \emph{First level}, \emph{Second level} with the definition (D) and one example (E) of each category in the column \emph{Definition and example}.
\figref{so-quora-full-dimension-categories} reports all levels of the second dimension according to the source. For instance, the questions about setting up tools (\emph{Tool setup}), or asking for various features  (\emph{Asking for features})  under \emph{IDE \& editors} are not found on Quora.
Similarly, the majority of the questions about documentation tools (\emph{Tools}) are asked on \SO whereas the general questions about comments (\emph{Commenting high levels}) are often on Quora.

\vspace{1mm}
\noindent\fbox{%
\parbox{\linewidth}{%
\fontsize{9}{10}\selectfont
\textbf{Finding 4.2.4} 
Developers often ask about the syntax to write function (method) comments compared to other kinds of comments (class, package), showing the trend of increasing effort towards API documentation.
Another frequently asked question on \SO concerns the conventions to add different kinds of information to code comments, such as code examples, media, or custom tags, indicating developers' interest in embedding various information in comments. 
}%
}

\begin{figure*}[ht] 
    \vspace{-3mm}
\includegraphics[width=\linewidth]{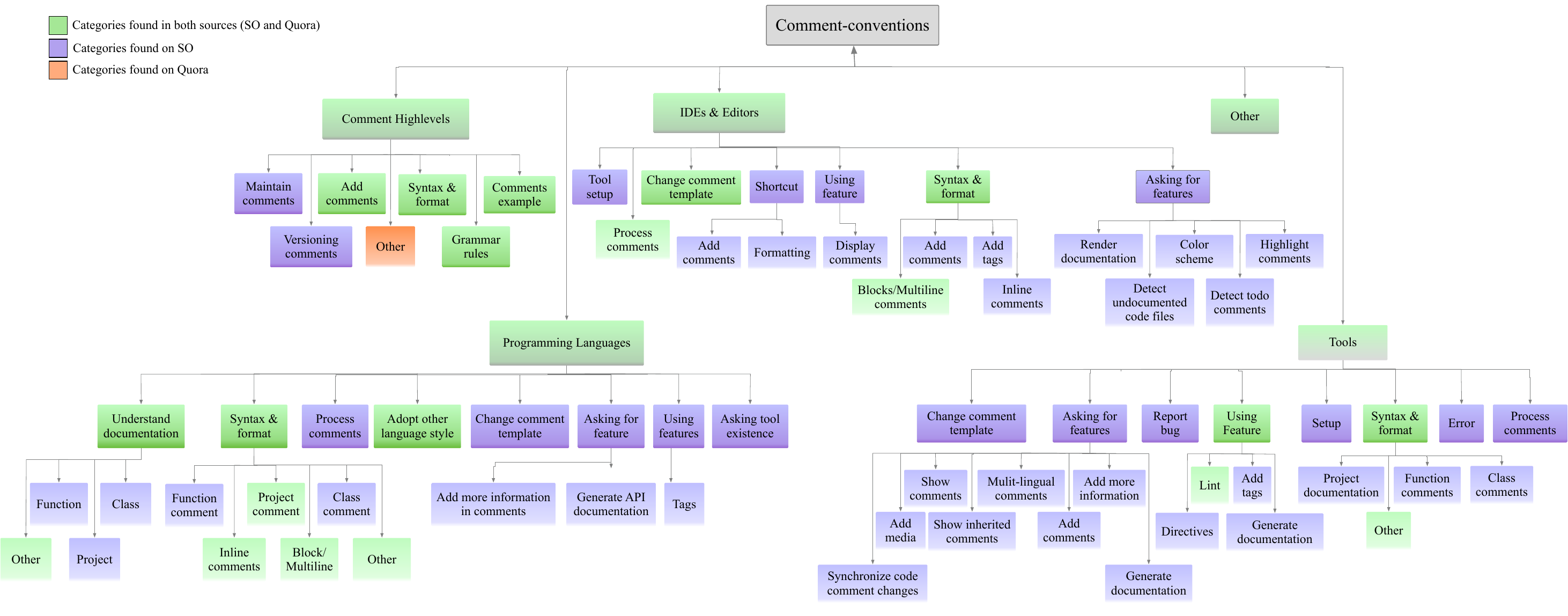}
\caption{Taxonomy of second dimension on \SO and Quora}
\figlabel{so-quora-full-dimension-categories}
    \vspace{-3mm}
\end{figure*}

\vspace{1mm}
\noindent\fbox{%
\parbox{\linewidth}{%
		\fontsize{9}{10}\selectfont
\textbf{Finding 4.2.5} 
Apart from questions related to comment syntax and features, developers ask about adopting commenting styles from other programming languages, modifying comment templates, understanding comments, and processing comments for various purposes.
}%
}
\vspace{1mm}

\subsection{Recommended Comment Convention (\textbf{RQ$_4$})} 
\begin{table}[htp]
\centering
\tiny
\caption{Code comment conventions recommended by developers}
\tablabel{comment-conventions}
%\begin{adjustbox}{width=0.8\linewidth}
		     \begin{center}
       \resizebox{0.80\linewidth}{!}{
       \scriptsize  
\begin{tabular}{p{0.08\linewidth}p{0.90\linewidth}}
\hline 
\textbf{Topic} & \textbf{Recommendation} \\
\hline
Grammar 
& [.net] long inline comments should start with a capital letter and end with a period.\\
& [.net] long inline comments should be written as a complete English sentence (with subject, verb, object).\\
& [general] check your coding style guidelines to verify how to write plural objects in the comments, for example, Things(s) or Things. \\
& [general] Do not mark the code section with an inline comment to highlight the modified code section, version control system keep track of code changes.\\
& [python] use backslash escape whitespace to use punctuations like apostrophe symbol in docstring.\\
& [python] Use `truthy' and `falsy' words to denote boolean values `True' and `False' respectively.\\
& [general] Do not write filler words such as `please' and `thank you', nor swearing words in the comments.\\
& [general] Remove TODO comments when you finish the task.\\
\hline
Language 
& [general] Comments should explain why and not how.\\
& [general] Use correct notation to write block or multiline comments.\\
& [general] Position your inline comments (about variable declaration) above the variable declaration to remain consistent with method comment conventions.\\
& [general] Do not write nested comments in the code.\\
& [general] Use different tags to categorize the information in the comments.\\
& [general] Do not use multiple single line comments instead of multi-line comments.\\
& [general] Do not document file specifications in the code comments rather document them in the design specs.\\
& [general] Use a consistent style such as \emph{`variable'} or \emph{$<$variable$>$} to differentiate the code variable names in the inline comments.\\
& [Java] Implementation notes about the class should be mentioned before the class definition rather than inside the class.\\
& [Java] To denote a method (\emph{someMethod() of the class ClassA }) in the comments, use the template \emph{the $<$someMethod$>$ method from the $<$ClassA$>$ class} instead of \emph{ClassA.someMethod()}.\\   
& [.net] Document `this' parameter of an extension method by describing the need of `this' object and its value.\\
& [javascript] \emph{Limitation:} Currently, there is no existing standard to document AJAX calls of javascript in PhpDoc style comments.\\
& [php] Use `-$>$' symbol to reference instance/object method rather than `::' in the method comments.\\
& [sql] Use the same documentation style for SQL objects as you are using for other code.\\
& [groovy] Limitation: there is no standard way to document properties of a dynamic map in Javadoc like JSDoc' @typedef.\\
\hline
Tool 
& [PhpDoc,JsDoc] Do not put implementation details of a public API in the API documentation comments, rather put them in inline comments inside the method.\\
& [JSDoc] Mention the class name in description to denote the instance of the class.\\
& [ghostDoc] Create your default comment template using c\# snippets.\\
& [JavaDoc] \emph{Limitation:} Currently it is not possible to generate documentation of an API in multi-languages (in addition to English) with the same source code.\\
& [JavaDoc]  \emph{Limitation:} The tag @value support fields having literal values. JavaDoc and IntelliJ IDEA do not support fetching value from an external file using @value tag in Javadocs. \\
& [Javadoc] Write annotations after the method javadoc, before the method definition.\\
& [Doxygen]  Use @copydoc tag to reuse the documentation from other entities.\\
& [Doxygen] \emph{Limitation:} Currently it is not possible to generate documentation of an API for different readers such as dev and users.\\
& [Doxygen] Use @verbatim / @endverbatim to document console input and output.\\
& [Roxygen] Limitation: Not possible to override docstrings so the parent docstring is used when inheriting a class.\\
& [PhpDoc] \emph{Limitation:} Currently it is not supported to document array details in the return type of a method. \\
& [PhpDoc] \emph{Limitation:} Currently, using @value tag or any similar tag to refer to the value of a field is not supported in PhpDoc, so developers should use @var tag instead. \\
& [Phpdoc] Use class aliases in import statement to write short name in docblock.\\
& [Sphinx] Limitation: It can't create sections for each class. Add yourself the sections in the .rst file. \\
\hline
\end{tabular}
}
\end{center}
%\end{adjustbox}
\vspace{-3mm}
\end{table} 
There are various syntactic and semantic commenting guidelines mentioned in the style guides, and developers are often confronted with several conventions, or are unable to find any for a specific purpose.
%For instance,  the syntactic guideline ``\emph{Use 3rd person (descriptive) not 2nd person} in writing comments'' in Java.
We collect various comment conventions recommended by developers in their answers on \SO and Quora in \tabref{comment-conventions}.
For example, a developer asks \emph{Should .net comments start with a capital letter and end with a period?}\SOid{2909241},
concerning grammar rules in the comments.
The accepted answer affirms the convention and describes how it helps to improve readability.
We, therefore, constructed the recommendation \emph{[.net] long inline comments should start with a capital letter and end with a period}.
In some answers, developers describe it as a limitation, we included \emph{Limitation} for such answers.
For each recommendation, we indicate whether it is specific to a programming language, a tool, an IDE, or is instead a general recommendation, using tags such as ``[Java], [Doxygen], [visual studio],[general]'' respectively.
It is important to note that we did not verify how widely the recommendations are adopted in the commenting style guidelines or projects, or how well they are supported by current documentation checker tools (or style checkers). 
This is a future direction for this work.
On the positive side, it represents an initial starting point to collect various comment conventions confirmed by developers.
We argue that it can also help researchers in conducting the studies to assess the relative importance of comment conventions or help tool developers in deciding which recommendation they should include in their tools to address frequent concerns of developers.
%These can guide developers in learning the best practices and in quickly reviewing their comments. 
%Additionally, these recommendations can help tool builders in supporting the most widely-adopted comment conventions in their tools.
%We proposed a list of recommendation given by experts about comment conventions which can help developers in quickly reviewing their comments, to assess adherence of comments to the coding standards, and to recommend best practices to developers gathered from \SO posts.

%===============================================================================
%=============================== Discussion =======================================
%===============================================================================

\section{Discussion and Implication}
\seclabel{discussion}
%In this section, we discuss developer concerns regarding code comments and discuss the main implications of our work.
\textbf{On Writing Comments.} Although, various coding style guidelines provide conventions to write comments, our results showed that \SO developers seek help in writing correct syntax of various comment types (class/function/package comments highlighted in \figref{syntax-format}), in adding specific information in comments, or formatting comments.
Typical types of questions are \emph{What is the preferred way of notating methods in comments?}~\SOid{982307}, and \emph{Indentation of inline comments in Python}~\SOid{56076686}, or \emph{indentation of commented code}~\SOid{19275316}.
This indicates the need of improving the commenting guideline and assuring their findability to developers.
Tomasottir \etal showed in their interview study that developers use linters to maintain code consistency
%to save discussion time, to speed up the code review process, 
and to learn about the programming language~\cite{Toma17a}.
%, especially in the case of novice developers~\cite{Toma17a}.
By configuring linters early in a project, developers can use them similarly to learn the correct syntax to write and format comments according to a particular style guideline.
However, due to their support to multiple languages, assisting developers in language-specific conventions, or customizing comments to add more information would still require further effort.

\textbf{On Coding Style Guidelines.}
Organizing the information in the comments is another concern highlighted in the study, for example, \emph{how to differentiate the variables within a comment}~\SOid{2989522}~\SOid{47089022}, \emph{where to put class implementation details} (in the class comment or in inline comments)~\SOid{35957906},
and \emph{which tag to use for a particular type of information}~\SOid{21823716}.
We also found developer concerns regarding grammar rules and word usage in all the sources we analyzed (\SO~\SOid{2909241}, and Quora~\cite{Quora:GR}).
Although various style guidelines propose comment conventions, there are still many aspects of comments for which either the conventions are not proposed or developers are unable to locate them.
Developers commonly ask questions, such as \emph{Any coding/commenting standards you use when modifying code?}~\SOid{779025} on \SO and \emph{Why is there no standard for coding style in GNU R?} on Quora.
There is therefore a need to cover detailed aspects of comments in the coding style guidelines to help developers write high-quality comments.

\textbf{On the Impact of Comment Conventions.}  Various commenting conventions are presented in the style guides to support consistent and readable comments.
However, extracting these conventions automatically from style guidelines and customizing them according to the project requirements is still a challenge and not explored much.
Additionally, which of these conventions play a more important role for comment comprehension and which ones do not, is not yet explored.
Binkley \etal evaluated the impact of identifier conventions on code comprehension, but the conventions were limited to identifiers~\cite{Bink13b}.
Smit \etal identified the relative importance of 71 code conventions, but the majority of the comment conventions were limited to detecting missing documentation comments~\cite{Smit11a}.
Therefore, assessing the impact and importance of comment conventions depending on a specific domain and project, and on various development tasks appears to be another potential direction.

\textbf{Tools to Assess Comment Quality.}
%Recently various research studies have investigated tools for automated generation of comments, automatic summarization of comments, detection of bad comments, identification of information embedded in comments, and the quality assessment of comments~\cite{Rato17a,Pasc17a,Kham10a}.
Our results show that developers are interested in various automated strategies, such as automatic generation of comments, detection of bad comments, identification of information embedded in comments, and the quality assessment of comments, lack tools that can be integrated into their IDE, especially to verify the comment style automatically~\SOid{14384136}.
However,  a limited set of documentation tools support comment quality assessment or adherence of comment to the commenting conventions.
For example, current style checker tools, such as Checkstyle, RuboCop and pydocstyle provide support for formatting conventions but lack support for comprehensive checks for grammar rules and content.\smallfootnote{[\url{https://checkstyle.org/checks.html}], [\url{https://rubocop.org/}], [\url{http://www.pydocstyle.org/}]}
There is a need to survey current automated style checker tools.
Additionally, some languages with advanced style checkers don't support comment checkers at all, such as OCLint for Objective-C, and Ktlint for Kotlin, Smalltalk.\smallfootnote{[\url{http://oclint.org/}], [\url{https://github.com/pinterest/ktlint}]}
We found instances of developers asking about the existence of such tools~\SOid{8834991} in \figref{asking-for-feature} and \emph{Asking tool existence} in Table III in the online appendix~\cite{Scam21a}.
Therefore, more tool support is needed to help developers in verifying the high-quality of comments.

%===============================================================================
%=============================== Threats to Validity ===================================
%===============================================================================

\section{Threats to Validity}
\label{sec:threats-to-validity}

\textbf{Threats to construct validity} concern the relationship between theory and experimentation.
In our study, they mainly relate to potential imprecision in our measurements. 
To mitigate potential bias in the selection of developer discussions on \SO, we relied on \SO tags to perform initial filtering. 
However, it is possible that this tag-based filtering approach misses some relevant posts concerning comment convention practices and topics.
%, because the tags used did not refer directly comment convention practices and topics. 
We therefore investigated the co-appearing tags to find similar relevant tags. 
Aghajani \etal studied software documentation-related posts, including code comments on \SO and other sources~\cite{Agha19a}.
We extracted their documentation-related tags from the given replication package and compared them to our tags ($It$) to verify if we missed any.
%We also validated our tags with the work of Aghajani \etal on software documentation\cite{Agha19a}.
On Quora, we mapped the selected \SO tags as keywords and searched these keywords on Quora search interface.
%The relevance of the topic was assessed by one of the authors.  
To avoid eventual biased in this manual process, we also adopted LDA,
to investigate high-level topics emerging in the \SO and Quora posts. Thus, a mix of qualitative and quantitative analysis was performed to minimize potential bias in our investigation, providing insights and direction into the automated extraction of relevant topics.

\textbf{Threats to internal validity} concern confounding factors, internal to the study, that can affect its results.
In our study, they mainly affect the protocol used to build the taxonomy, which could directly or indirectly influence our results. 
To limit this threat, we used different strategies to avoid any subjectivity in our results. 
Specifically, all posts were validated by at least two reviewers and, in case of disagreement,  a third reviewer participated in the discussion to reach a consensus. 
Thus, for the definition of the taxonomy, we applied multiple iterations, involving different authors of this work.

\textbf{Threats to conclusion validity} concern the relationship
between theory and outcome. In our study, they mainly relate to the extent to which the produced taxonomy can be considered exhaustive. 
To limit this threat, we focused on more than one source of information (SO and Quora), so that the resulting taxonomy has a higher likelihood to be composed of an exhaustive list of elements (\ie comment convention topics). 

\textbf{Threats to external validity} concern the generalizability
of our findings. 
These are mainly due to the choice of \SO and Quora as the main sources. 
\SO and Quora are widely used for development discussions to date,
although specific forums, such as DZone and Reddit could be considered for future works.\smallfootnote{[\url{https://dzone.com/articles/my-commentary-on-code-comments}], [\url{https://www.reddit.com}]}
Moreover, besides all written sources of information, we are aware that there is still a portion of the developer communication taking place about these topics that are not traceable.
Thus, further studies are needed to verify the generalizability of our findings.

%===============================================================================
%=============================== Related Work ===================================
%===============================================================================
\section{Related Work}
\label{sec:related-work}

Studying developer activities related to various development tasks from various sources can guide the development of tools that help developers to find the desired information more easily.
Therefore, researchers have recently focused on leveraging the useful content of these sources \ie Git, CVS~\cite{Chen16a}, archived communications 
%\eg \ml~\cite{Shar11a,Shar08a}, execution logs~\cite{Godf08a}, Newsgroups~\cite{Hou05a}, 
online forums and CQA (Community Question Answer) sites~\cite{Alla13a,Baru14a,Rose16a,Yang16a,Krug19a,Agha19a} 
%~\cite{Alla13a,Baru14a,Baja14a,Treu11a,Yang16a,Wu15a} 
to comprehend developers information needs.
\SO is one of the more popular platforms that researchers have studied to capture developers questions about trends and technologies~\cite{Alla13a}, security-related issues~\cite{Yang16a}, and documentation issues \etc~\cite{Agha19a}.
Recently researchers have started investigating Quora to get more insight into developer communities~\cite{Krug19a}, \eg finding and predicting popularity of the topics~\cite{Wang13c,Krug19a}, finding answerability of the questions~\cite{Mait17a}, detecting experts on specific topics~\cite{Pati16a,Nesh17a,Geer16s}, or analyzing anonymous answers~\cite{Math19a}
Our study is first to investigate this platform for code comments.

Research shows that interesting insights can be obtained from combining these sources~\cite{Bavo16a,Agha19a}.
Aghajani \etal studied documentation issues on \SO, Github, and \ml~\cite{Agha19a}.
They reported a taxonomy of documentation issues developers face.
However, they do not focus on the style issues of the code comments.
Our study focuses on the all aspects of code comments \ie the content and style aspect of the code comments.
In a previous study, Barua \etal found coding style/practice among the top share on \SO~\cite{Baru14a}.
They considered the topic among common English language topics instead of a technical category due to usage of generic words in this topic. 
As their focus was on technical categories, they did not explore the coding style questions further.
Our study complements their work by exploring the specific aspects of coding style, focusing on comment conventions.

%===============================================================================
%=============================== CONCLUSION ===================================
%===============================================================================
\section{Conclusions}
\label{sec:conclusion}
In this study, we investigated commenting practices discussions occurring in \SO, and Quora.
We first performed automated analysis (LDA) on extracted discussions and then complemented it with a more in-depth manual analysis on the selected sample set.
From the manual analysis, we derived a two-dimensional taxonomy.
The first dimension of the taxonomy focuses on the question types, while the second dimension focuses on five types of first-level concerns and 20 types of second-level concerns developers express.
%First level concerns present the leading topics developers talk about, such as an IDE, an editor, a documentation tool, or a programming language.
%Second level concerns show the specific concerns in comment conventions about writing the correct syntax, asking for a particular feature, understanding a part of the documentation (in first-level categories).
We qualitatively discussed our insights, and presented implications for developers, researchers and tool designers to satisfy developer information needs regarding commenting practices.
We provide the data used in our study, including the validated data and the detailed taxonomy, in the replication package~\cite{Scam21a}.
%We also presented a list of recommendations given by experts on \SO and Quora to help developers in verifying their comments conventions.
%In particular, this can help developers in quickly checking the quality of their comments, code reviewers to spot the issues in the comments, and project maintainers to verify the consistency in comments.
In the future, we plan (i) to verify the completeness and relevance of gathered comment conventions, (ii) to survey practitioners and tool designers to learn which rules are more important than others, and (iii) to explore ways to improve and assess tool support in this direction. 
This investigation will help us to see which comment convention affects them most, and during which specific development activity.

\section*{Acknowledgments}
We gratefully acknowledge the financial support of the Swiss National Science Foundation for the project ``Agile Software Assistance'' (SNSF project No.\ 200020-181973, Feb.\ 1, 2019 - April 30, 2022).

\bibliographystyle{IEEEtran}
\bibliography{commentconvention}

\end{document}

\title{Appendix of the Paper {``What do Developers Discuss about Code Comments?''}}

\author{Authors details omitted for double-blind reviewing}

\maketitle

\appendix

\section{Detailed Taxonomy}
\tabref{information-type} shows the hierarchy of categories constructed in the second dimension \ie \emph{First level}, \emph{Second level} with the definition (D) and one example (E) of each category in the column \emph{Definition and example}.

\begin{table}[htp]
		\centering
		\tiny
		\caption{Type of Information developers seek on \SO and Quora}
		\tablabel{information-type}
		\begin{adjustbox}{width=\linewidth}  
		\begin{tabular}{p{0.10\linewidth}p{0.17\linewidth}p{0.72\linewidth}}
		\hline 
		\textbf{First level} & \textbf{Second level} & \textbf{Definition and Example} \\
		\hline
		
		Commenting High levels & Add comments
		& D: general questions about adding comments without mentioning a specific programming language, tool or IDE \\
		& & E: \emph{whats the most professional and informative way of commenting code?} \\
		
		& Versioning comments 
		& D: questioner asks about best practices for comments in code versioning tools like git or svn.\\
		& & E: \emph{how to formulate a commit message?} \\
		
		& Comments example 
		& D: questioner asks for specific examples (funny, helpful, silly) of code comments they have seen.\\
		& & E: \emph{What's the least useful comment you've ever seen?}  \\
		
		& Grammar rules & D: ask about following grammar rules in writing comments.\\
		& & E: \emph{Should .net comments start with a capital letter and end with a period?} \\
		
		& Maintain comments
		& D: general questions about maintaining comments over time.  \\
		& & E: \emph{Maintenance commenting}\\
		
		& Other
		& D: general conceptual questions about code comments \\
		& & E: \emph{What's a good comment/code ratio?}\\
		
		& Syntax \& format 
		& D: general questions about syntax and format of comments irrespective of a development environment \\
		& & E: \emph{Documentation style: how do you differentiate variable names from the rest of the text within a comment?}\\
		\noalign{\smallskip}\hline
		 Languages & Adopt other language style
		 & D:  questions about adopting the commenting style of another programming language\\
		 & & E: \emph{Is it a bad practice to use C-style comments in C++ code?}\\
		
		 & Asking for Feature 
		 & D: questions regarding whether a feature is supported or not and if not, then how a problem can be solved in the language \\
		 & & E: \emph{How do I put code examples in .NET XML comments?}  \\
		
		 & Asking tool existence 
		 & D: Users ask whether there is a tool for a particular programming language to document code \\
		 & & E: \emph{Is there any specific tool that is used by underscore authors to generate documentation in javascript?}  \\
		 
		 & Change comment template 
		 & D: questions about modification in the template of a specific programming language or environment.\\
		 & & E: \emph{How to propose some revision to RFC(s) for JSON to account for comments?} \\
		
		 & Process comments 
		 & D: questions about processing comments of a particular programming language. Processing includes striping, removing, extracting, cleaning comments \\
		 & & E: \emph{Remove comment blocks bounded by ``\#|...|\#'' in textfile - python}  \\
		
		 & Syntax \& Format 
		 & D: questions about the syntax of comments in a specific programming language\\
		 & & E: \emph{Commenting in c++?} \\
		
		 & Understand documentation 
		 & D: questioner face difficulties with understanding code documentation of function, class, or project\\
		 & & E: \emph{How to interpret cryptic Java class documentation?} \\
		 
		 & Using features 
		 & D: user is aware about the feature but does not know how to use the feature \\
		 & & E: \emph{What are these tags @ivar @param and @type in python docstring?}  \\
		
		 \noalign{\smallskip}\hline
		 Tools & Asking for features 
		 & D: question regarding whether a feature is supported or not, or how a problem can be solved with the tool.\\
		 & & E: \emph{How to properly write cross-references to external documentation with intersphinx?}  \\
		  
		 & Change comment template
		 & D: questions about modification in the comment template provided by the tool.\\
		 & & E: \emph{Qt-style documentation using Doxygen?} \\
		 
		& Error
		 & D: questioner needs help with some error or warning received through the tool while writing documentation\\
		 & & E: \emph{Stylecop doesn't understand $<$inheritdoc$>$}  \\
		
		 & Process comments 
		 & D: questions about processing comments in a tool. Processing includes striping, removing, extracting, cleaning comments \\
		 & & E: \emph{Gradle groovy how to keep comments and all formats in XML parser}  \\
		 
		 & Report Bug
		 & D: questioner reports a (potential) bug.\\
		 & & E: \emph{Doxygen C\# XML comments and generics do not generate links in HTML output?} \\
		 
		 & Setup
		 & D: questioner ask about the configuring the tool\\
		 & & E: \emph{How can I configure GhostDoc to generate comments for attributes on properties?} \\
		 
		 & Syntax \& Format 
		 & D: questioner asks ways to document specific code elements such as class, methods or parts of code in a documentation tool.\\
		 & & E: \emph{How do I refer to classes and methods in other files my project with Sphinx?} \\
		 
		 & Using Feature 
		 & D: question regarding how to use a certain feature of the tool.\\
		 & & E:  \emph{How to use @value tag in javadoc?}\\
		 
		 \noalign{\smallskip}\hline
		
		 IDEs \& Editors & Asking for Feature 
		 & D: questions regarding whether a feature is supported or how a problem can be solved in the environment \\
		 & & E: \emph{Android - (Android Studio) - Create similar to TODO, but different COLOR AND NAME?}  \\
		
		 & Change comment template 
		 & D: questions about modification in the template in the IDE or editor.\\
		 & & E: \emph{Add different default and custom tags to Visual Studio XML Documentation} \\
		
		 & Process comments 
		 & D: questions about processing comments in an IDE or an editor. Processing includes striping, removing, extracting, cleaning comments \\
		 & & E: \emph{How do I get rid of XXX Auto-generated method stub?}  \\
		
		  & Shortcut  
		 & D: question regarding how to achieve certain functionality with a shortcut (keyboard)\\
		 & & E: \emph{Finding Shortcuts in Aptana Studio 3.0 to Comment Code} \\
		
		 & Syntax \& Format 
		 & D: questions about the syntax of comments in an IDE or editors\\
		 & & E: \emph{How to comment SQL statements in Notepad++?} \\
		
		 & Tool setup 
		 & D: setup a documentation tool in a particular IDE or editor \\
		 & & E: \emph{How to generate javadoc using ubuntu + eclipse to my project}  \\
		
		 & Using features 
		 & D: user is aware about the feature but does not know how to use the feature \\
		 & & E: \emph{How can I get Xcode to show my documentation comments during option-hover?}  \\
		 \noalign{\smallskip}\hline
		Other &
		& D: the questions not belonging to any above category\\
		& & E: \emph{Why does Godot receive praise for its fantastic documentation and ease of use from a coding perspective?}\\
		\hline
		\end{tabular}
		\end{adjustbox}
	\end{table}

\section{Our study reproducibility}
\seclabel{reproducibility-case-study}

Gonz{\'a}lez \etal identified the reproducibility aspects characterizing empirical software engineering studies: {Data Extraction}, {Data Preprocessing}, {Dataset Availability}~\cite{Gonz12a}.
To address these reproducibility concerns in our study, we used the tool named \emph{\makar}~\cite{Maka20a}.
%The tool is made available as a docker image to facilitate its usage on different platforms and operating systems.
%As follow, we explain how \emph{\makar} allows us to address each of the identified \textit{reproducibility dimension}.

\textbf{Data Extraction.}
%\label{subsec:case-study-data-extraction}
%To extract data from selected sources such as \SO, Quora, and \ML, we used different data extraction methods.
For \SO, we used the public API provided by Stack Exchange platform whereas for Quora, we scraped the source by developing our own crawlers. 
%Detailed information concerning the extracted fields from each source is reported in \autoref{subsec:data-collection}.
\textbf{Data Preprocessing.}
%\label{subsec:case-study-data-preparation}
%Data preprocessing in our study was only required to prepare the data for LDA analysis (RQ$_1$)
The data from Stack Overflow contains HTML, code snippets, links and natural language text.
To obtain meaningful results from LDA analysis, it was necessary to clean the data.
% with \emph{text cleaning} and \emph{language cleaning} steps.
%To perform the manual analysis and to construct the taxonomy, we did not require any pre-processing steps on the data.
\figref{case-study-preprocessing} illustrates the preprocessing steps performed by \makar.
The \emph{Transformation} describes various built-in transformations of \makar and \emph{Attributes} shows the list of selected fields (described in the paper) from the sources.
The transformations have been set up so that each step produced a new attribute on the data record, allowing us not to lose any information, and retrace every change we made to the data.
At the beginning of the case study, it was unclear which combination of question and answer attributes leads to the best result. 
In this phase, the flexible approach to preprocessing found in \makar helped us to work with our data and try different options efficiently.
\begin{figure}[h]
\centering
\includegraphics[width=\linewidth]{plots/05_preprocessing}
\caption{All preprocessing steps with the used transformation in \emph{\makar}}
\figlabel{case-study-preprocessing}
\end{figure}

All preprocessing steps are performed with \makar using its built-in transformations:
\begin{itemize}
	\item \emph{Code} is removed with the \textit{extract\_code} transformation. 
	This transformation extracts \verb|$<$code$>$| tags from the given attributes and saves it to a new attribute.
	For the case study, we did not need the extracted code snippets for further analysis, but this step would allow for easy access to the extracted snippets.
	%This transformation was applied only to the question body as the question title does not contain any code.
	\item \emph{HTML} tags are removed with the \textit{strip\_html} transformation.
	This transformation removes all HTML tags from the given attribute. 
	Only the tags are removed, not the content of tags.
	%This transformation was  applied only to the question body as the question titles do not contain any HTML tags.
	\item \emph{Punctuation Removal} is done with the \textit{string\_replace} transformation. 
	This transformation accepts a string or regular expression as search parameter and a string as replacement parameter.
	For this preprocessing step, we replaced \verb|/[.!?\\*\-\_&#$+\%()\[\]{}]/| with the empty string, effectively removing punctuation and special characters.
	%This transformation was applied to each question body and title.
	\item \emph{Stop Word Removal} is done with the dedicated \textit{remove\_stopwords} transformation.
	This transformation removes all stop words from the \textit{Snowball English Stop Word List}\smallfootnote{\url{http://snowball.tartarus.org/algorithms/english/stop.txt}}, effectively removing very common words that would not contribute to usable results for topic modeling.
	%This transformation was applied to question body and title.
	\item \emph{Word Stemming} is done with the \textit{word\_stemming} transformation.
	This transformation applies the \textit{Snowball Stemming Algorithm}\smallfootnote{\url{https://snowballstem.org/algorithms/porter/stemmer.html}}, which is an implementation of the \textit{Porter Stemming Algorithm}\smallfootnote{\url{https://tartarus.org/martin/PorterStemmer/}}, to the given attribute.
	%This transformation was applied to question body and title.
\end{itemize}

\textbf{Dataset Availability}
%\label{subsec:dataset-availability}
To make our dataset available, we provide a replication package containing raw data, validated data and various other supporting material~\cite{Scam21a}.
		
\bibliographystyle{IEEEtran}
\bibliography{scg,references}